\documentclass[pre,twocolumn,showpacs]{revtex4}
\usepackage{epsf}
\usepackage{epsfig}
\begin{document}
\title{
Aging Continuous Time Random Walks.
}
\author{Eli Barkai} 
\affiliation{Department of Chemistry and Biochemistry, Notre Dame University, Notre Dame, IN 46556.}
\email{jbarkai@nd.edu}
\author{Yuan-Chung Cheng}
\affiliation{Department of Chemistry,
Massachusetts Institute of Technology,
Cambridge, MA 02139.
}
\date{\today}

\begin{abstract}

We investigate aging continuous time random walks (ACTRW),
introduced by Monthus and Bouchaud
[{\em J. Phys. A} {\bf 29}, 3847 (1996)].
Statistical behaviors of the  displacement of the random walker
${\bf r}={\bf r}(t) - {\bf r}(0)$ in the time interval
$(0,t)$ are obtained,
after aging the random walk in the time interval $(-t_a,0)$.
In ACTRW formalism, the Green function
$P({\bf r}, t_a , t)$ depends on the age of the random walk
$t_a$ and the forward time $t$.
We derive a generalized Montroll--Weiss equation, which yields
an exact expression for the Fourier double Laplace transform
of the ACTRW Green function.
Asymptotic long times
$t_a$ and $t$ behaviors of the Green function
are investigated in detail.
In the limit of $t\gg t_a$, we recover the standard non-equilibrium
CTRW behaviors,
while the important regimes
$t\ll t_a$ and $t \simeq t_a$ exhibit interesting aging effects.
Convergence of the ACTRW results 
towards the
CTRW behavior,
becomes extremely slow when the diffusion exponent becomes small.
In the context of biased ACTRW, we investigate an aging Einstein relation.
We briefly discuss aging in Scher-Montroll type of transport
in disordered materials.

\end{abstract}

\maketitle

\section{Introduction}

 Diffusion and relaxation in strongly disordered systems exhibits in many cases
anomalous behaviors \cite{Bouch1,review,Havlin}. 
For example the diffusion of a test particle
may become anomalous, namely the mean square displacement
behaves like $\langle r^2 \rangle \sim t^{\alpha}$ and $\alpha \ne 1$.
A random walk framework, widely applied to describe anomalous
diffusion is the continuous time random walk
(CTRW) \cite{Weiss1}.
CTRWs are used to model many physical and chemical processes, for example:
charge transport in disordered systems \cite{SM},
protein folding dynamics \cite{Wolynes,Plotkin,Wang},
transport in low dimensional chaotic systems \cite{Klafter1,Swinney,SZ},
anomalous diffusion in a  metallic supercooled liquid
\cite{Gezelter}, the chemical reaction of
CO binding to myoglobin \cite{Jaeyoung},
 and blinking behavior of single quantum dots
\cite{Jung,Brokmann}.

Anomalous diffusion processes may 
exhibit aging \cite{VF,Parisi,Fisher,Laloux,Sokolov}, where
vaguely speaking the age of the process controls
the statistical properties of the random walk. 
Aging in diffusion processes yields an interesting insightful perspective 
on dynamics in disordered medium, and more generally is used as
a tool to probe complex systems such as spin glass,
Anderson insulator, and colloidal suspensions
(see \cite{Hike} for a brief review). 
Monthus and Bouchaud  \cite{Month} introduced
a CTRW framework, which exhibits aging behaviors.
We call this generalized CTRW, aging continuous time random
walk (ACTRW). In this paper, we investigate properties
of biased and non-biased ACTRWs in detail,
in particular an exact expression for the Fourier
double Laplace transform of the Green function is obtained,
and asymptotic behaviors are investigated.

 ACTRW might yield a phenomenological description of aging 
self diffusion dynamics in glasses \cite{Month},
below the glass transition temperature
$T_c$ and then $\alpha=T/T_c$ is temperature
dependent 
(and see \cite{Rinn,Sollich,Evans,Heuer,Reichmann} for
related work).
Recently  
\cite{Barkaiage} showed that ACTRW describes dynamics of an 
intermittent chaotic system. More generally, we expect that any random walk
described by a CTRW process will exhibit aging, 
provided that aging initial condition discussed in the manuscript
are satisfied. 
We note that other stochastic approaches to
aging dynamics are based on a non-linear diffusion equation \cite{Stariolo},
and a generalized Langevin equation \cite{Pot}.

 This paper is organized as follows.
In Sec. 
(\ref{Secctvs}) CTRW and ACTRW are introduced.
 We then derive an exact expression for the
Fourier double Laplace  transform of the ACTRW Green function
$P({\bf r},t_a,t)$, 
thus generalizing the Montroll-Weiss equation 
(\ref{eq01})
to include the
effect of the age of the random walk (see Sec. \ref{secGMWE}). 
Our result is based in part on recent work of Gordeche and Luck \cite{GL}
who investigated renewal theory for
processes with power law waiting time distributions.
In Sec. \ref{SecAB} we derive asymptotic behaviors of the Green function
$P({\bf r},t_a,t)$ which are analyzed in detail.
We then consider biased ACTRWs (Sec. \ref{SecBi}) and discuss
an aging Einstein relation. Finally we briefly consider aging
in Scher-Montroll type of transport as a
possible application of this work.
Note that a small part of our results was reported in
\cite{Barkaiage}, in the context of aging in chaotic transport.

\section{CTRW and ACTRW}
\label{Secctvs}

 One of the best well known random walk models 
is the CTRW introduced by 
Montroll and Weiss \cite{MW}.
 It 
describes a large class of random walks, both normal and anomalous
and can be described as follows. Suppose a particle performs
a random walk in such a way that the individual jump ${\bf r}$ in
space is governed by a probability density function (PDF)
 $f({\bf r})$, and that all jump
vectors are independent and identically distributed.
The characteristic function of the position of the particle
relative to the origin after $n$ jumps is $f^n({\bf k})$,
where $f({\bf k})$ is the Fourier transform of $f({\bf r})$.
Unlike discrete time random walks, the CTRW describes a situation
where the waiting time $t$ between jumps is not a constant.
Rather,
the waiting time is
governed by the
PDF $\psi(t)$ and all waiting times
are mutually independent and identically distributed.
Thus, number of jumps $n$ is a random variable.

 Let $P_{MW}({\bf r}, t)$ be the Green function of the CTRW, the 
Montroll--Weiss
equation yields this function in Fourier--Laplace $({\bf k},u)$ space:
\begin{equation}
P_{MW}\left( {\bf k}, u\right) = { 1 - \psi(u) \over u} { 1 \over 1 - f\left( {\bf k} \right) \psi\left(u\right)}.
\label{eq01}
\end{equation}
All along this work we will use the convention that the arguments in the 
parenthesis define the space we are working in, thus $\psi(u)$
is the Laplace transform of $\psi(t)$. 
 Properties of 
$P_{MW}({\bf r}, t)$ based on the Fourier--Laplace inversion of
Eq. (\ref{eq01}) are well investigated, see \cite{Weiss1,Tran}
and Ref. therein. In particular, it is well known that asymptotic behavior
of $P_{MW}\left( {\bf r}, t\right)$ depends on the long time behavior
of $\psi(t)$. Two classes of processes are usually considered. The first
is the case when all moments of $\psi(t)$ are finite, the second class is
the case where $\psi(t)$ is moment-less, corresponding to a situation
where $\psi(t) \propto t^{-(1 + \alpha)}$ and $0<\alpha <1$.

An important assumption made in the derivation of Eq. (\ref{eq01})
is that the random walk begun at time $t=0$. More precisely,
it is assumed that the PDF of the first waiting time, i.e., the time
elapsing between start of the process at $t=0$ and the first
jump event is $\psi(t)$. 
Thus the Montroll-Weiss CTRW approach describes a particular
choice of initial conditions, called non-equilibrium initial
conditions. The limitation of CTRW theory to a very particular
choice of initial conditions, was an issue
for debates in the early seventies \cite{Tunaley}.

\begin{figure}[htb]
\epsfxsize=20\baselineskip
\centerline{\vbox{
           \epsfig{file=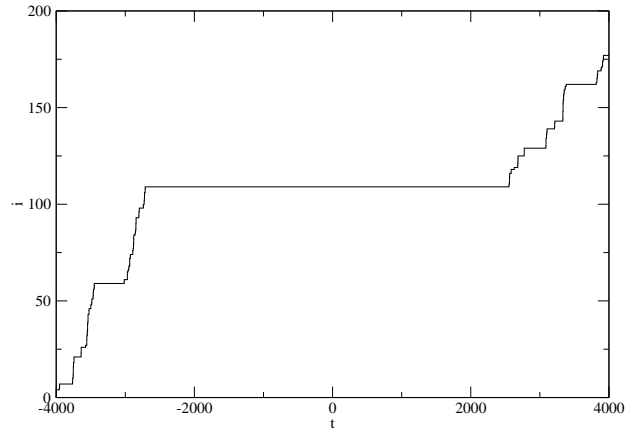, width=0.85\linewidth, angle=-90}  }}
\caption {
Number of jumps $i$ in a renewal process with $\psi(t)=\pi^{-1}t^{-1/2}(1+t)^{-1}$, i.e. $\alpha=1/2$. The process
 starts
at $t=-t_a$ with $t_a=4000$. Observation of the process
begins at time $t=0$. 
$t_1$ is the time elapsing between $t=0$ and first jump event
in the forward time interval $(0,t)$, in the Fig.  $t_1=2553$.
}
\label{fig00}
\end{figure}

 Monthus and Bouchaud \cite{Month} 
introduced a CTRW for an ongoing
process, 
where the random walk process is assumed to start at some time 
$t=-t_a$, long before start of observation at time $t=0$.
In Fig. \ref{fig00} a stochastic realization of number of jumps
in such a process is shown.
For such a random walk the Green function
is denoted with $P({\bf r}, t_a, t)$ 
and ${\bf r}$ is the displacement in the time interval $(0,t)$.
Using scaling analysis, \cite{Month} have investigated 
basic properties
of this random walk, mainly the behavior of the Fourier transform
of the Green function. 
For $\alpha<1$ the random walk
exhibits interesting aging effects, hence as mentioned
we call it
ACTRW. 

\section{ACTRW:- Generalized Montroll--Weiss Equation}
\label{secGMWE}

The ACTRW describes the following process, a particle
is trapped on the origin for time $t_1$, it then jumps
to ${\bf r_1}$, the particle is then
trapped on ${\bf r_1}$ for time $t_2$,
and then it jumps to a new location; the process is then renewed.
Thus, the ACTRW process
is characterized by a set
of waiting times $ \{ t_1, ... , t_n , ...\} $ and
displacements $\{ {\bf r_1}, \cdots , {\bf r_n} \cdots\} $.
The time elapsing between start of observation at
 $t=0$, and the first jump event 
is denoted by $t_1$. 
Here we denote the PDF of the first waiting time $t_1$ with
$h_{t_a}(t_1)$. 
In ACTRW the random walk
started at $t=-t_a$, before the start of observation at $t=0$,
therefore  
$h_{t_a}(t_1)$ depends on age of the process $t_a$.  
The
waiting times $\{ t_n \}$, with
 $n>1$ are independent and identically distributed
with a common probability density
$\psi(t)$. The jump length $\{ {\bf r_1}, \cdots , {\bf r_n} \cdots\} $,
are independent identically distributed random variables, described
by the probability density $f({\bf r})$. 

In contrast, in the Montroll--Weiss non-equilibrium CTRW, the age of the process
is zero $t_a=0$. 
And, for that case
$h_{t_a}(t_1) = \psi(t_1)$.  
 
 Recently, Gordeche and Luck \cite{GL} investigated statistical
properties of fractal renewal processes, among other things  
they obtain $h_{t_a}(t_1)$. 
 Let $h_s(u)$ be
the double Laplace transform of $h_{t_a}(t_1)$
\begin{equation}
h_s(u) = \int_0^{\infty} {\rm d} t_1 \int_0^{\infty} {\rm d} t_a 
h_{t_a} \left( t_1 \right) e^{- t_a s - t_1 u},
\label{eq03}
\end{equation}
then according to \cite{GL}
\begin{equation}
h_s(u) = {1 \over 1 - \psi(s)} {\psi(s) - \psi(u) \over u - s}.
\label{eq04}
\end{equation}
In Appendix A we re-derive Eq. (\ref{eq04}) using a method
which slightly differs than the one used in \cite{GL}.
 
Two types of behaviors are found for $h_{t_a}(t_1)$ \cite{GL}.
 The first case
corresponds to a situation when average waiting time 
$\langle t \rangle=\int_0^{\infty} t \psi(t) {\rm d} t$ 
is finite, and then in the long aging time limit one obtains
\begin{equation}
\lim_{t_a \to \infty} h_{t_a} \left( t_1 \right) = { \int_0^{t_1} \psi(t) {\rm d} t \over \langle t \rangle}.
\label{eq04a}
\end{equation}
This type of initial condition is called equilibrium initial
condition, it was investigated previously in the context 
of CTRW and related models
\cite{Tunaley,Barkai8,Shushin}.
Here we will mainly consider the second case corresponding to a
power law waiting time PDF
\begin{equation}
\psi(t) \propto t^{ - (1 + \alpha)} \ \ \mbox{with} \ \ 0<\alpha<1, 
\label{eqpprr}
\end{equation}
when $t$ is long. In Laplace $t \to u$ space Eq. (\ref{eqpprr})
reads
\begin{equation}
\psi(u) \sim 1 - A u^{\alpha},
\label{eqpprr1}
\end{equation}
where $u$ is small, and $A$ is a positive parameter \cite{remark}.
For example the one sided L\'evy PDFs whose
Laplace pair is $\psi(u)=\exp( - A u^{\alpha})$,
or $\psi(u) = 1/(1+ A u^{\alpha})$ discussed below, belong 
to the class of functions described by Eqs. (\ref{eqpprr},\ref{eqpprr1}).
In the
limit of long aging times, $t_a >> A^{1/\alpha}$, these
kind of probability densities yield 
\begin{equation}
h_{t_a}(t_1) \sim { \sin(\pi \alpha) \over \pi} { t_a^{\alpha} \over t_1^{\alpha}
\left( t_1 + t_a \right)}.
\label{eq05}
\end{equation}
Note that this expression is independent of the exact form of
$\psi(t)$, except for the exponent $\alpha$. When $\alpha \to 1$
the mass of the PDF $h_{t_a}(t_1 )$ is concentrated in the vicinity
of $t_1 \to 0$, as expected from a `normal process'.
Eq. (\ref{eq05})
shows that as age of the process becomes older, we have to
wait longer for first jumping event to occur. 
In a physical process, this may correspond to a particle in a disordered
system which searches for a local energy minima in time 
interval $(-t_a,0)$. In this case the longer the search takes place the
deeper the minima found, hence in statistical sense
 the release time becomes
longer as the process is older. 
In what follows, we will also use the double Laplace transform of Eq.
(\ref{eq05}):
\begin{equation}
h_s(u) \sim { u ^{\alpha} - s^{\alpha} \over s^{\alpha} ( u - s)}.
\label{eq06}
\end{equation}
This equation can be derived by  inserting the small $u$ and $s$
 expansion of $\psi(u)$ and $\psi(s)$ given in  Eq. (\ref{eqpprr1}), 
in Eq.
(\ref{eq04}).

 Let $P({\bf r}, t_a, t)$ be the Green function of the random walker,
where as mentioned 
\begin{equation}
{\bf r}\equiv {\bf r} \left( t \right) - {\bf r } \left( 0 \right)
\label{eqdef}
\end{equation}
 is the displacement in the
time interval $(0,t)$.
 Hence, clearly  initially ${\bf r} =0$ at time $t=0$.
Let $(i)$ $p_{n}(t_a,t)$ be the
probability of making $n$ steps in the time interval
$(0,t)$ and $(ii)$
$P({\bf k},s,u)$ be the double--Laplace
--Fourier transform $({\bf r} \to {\bf k}, t_a \to s, t \to u)$ of
$P({\bf r}, t_a, t)$. Then 
\begin{equation}
P({\bf k},s,u)= \sum_{n = 0}^{\infty} p_n(s,u) f^n\left({\bf k}\right),
\label{eq07}
\end{equation}
where $p_n(s,u)$ is the double Laplace transform of
$p_n(t_a,t)$. As mentioned, $f^n ({\bf k})$ in Eq. (\ref{eq07}) is
the characteristic function of a random walk with exactly
$n$ steps. Using the convolution theorem of Laplace transform
we obtain
\begin{equation}
p_n(s,u) = \left\{
\begin{array}{l l}
{1- s h_s(u) \over s u} \ & n=0 \\
\ & \  \\
h_s(u) \psi^{n-1}(u) {1 - \psi(u) \over u} \ & n \ge 1 .
\end{array}
\right.
\label{eq08}
\end{equation}
Hence inserting Eq. (\ref{eq08}) in Eq. (\ref{eq07}), using Eq. (\ref{eq04}),
and summing over $n$,
we find the exact result
\begin{equation}
P({\bf k},s,u) = { 1  \over  s u} + {\left[ \psi\left(u\right) - \psi\left(s \right)\right] \left[ 1 - f\left( {\bf k} \right)\right] \over u \left( u -s \right) \left[ 1 - \psi\left(s \right) \right] \left[ 1 - \psi\left(u \right) f \left( {\bf k } \right) \right] }.
\label{eq09}
\end{equation}
Eq. (\ref{eq09}) is a generalization of the Montroll--Weiss equation
(\ref{eq01})
for ACTRW.  
Note that $P\left( {\bf k}=0,s,u \right) = 1/(s u)$ as expected from 
the normalization condition.

 It is useful to rewrite Eq. (\ref{eq09}) in terms of the Montroll--Weiss
Eq. (\ref{eq01}), and $p_{0}(s,u)$ in the first line of Eq. (\ref{eq08}):
\begin{equation}
P({\bf k},s,u) = p_{0}(s,u) + h_s(u) f({\bf k}) P_{MW}({\bf k},u).
\label{eq10}
\end{equation}
The first term on the right hand side of this equation,
describes random walks where the particle does not leave the origin 
(i.e. $n=0$).
The second term describes random walks where number of steps is
greater than zero, it is given in terms of
a convolution of $h_s(u) f({\bf k})$ with the Montroll-Weiss
equation. This is expected since the only difference between
ACTRW
and the non-equilibrium CTRW,
is the first waiting time distribution.

 If the process is Poissonian, $\psi(t)=\exp( - t)$, the Green function
$P({\bf r},t,t_a)$ is independent of the age of the random walk
$t_a$. To show this we insert
\begin{equation}
\psi(u) = {1\over 1 + u}, \ \ \ \psi(s) ={1 \over 1 + s}
\label{eq10aa}
\end{equation}
in Eq. (\ref{eq09}) and  find
\begin{equation}
P\left( {\bf k}, s, u \right) = { 1 \over s} { 1 \over u + 1 - f ( {\bf  k} ) }.
\label{eq10bb}
\end{equation}
Inverting to the double time domain
\begin{equation}
 P\left( {\bf k }, t_a , t \right) =  e^{ - \left[1-f\left({\bf k }\right) \right]t }.
\label{eq10aaa}
\end{equation}
This result  is independent of $t_a$ as expected from a Markovian 
process. Assume that $f({\bf k }) = 1 - m_{\mu} |{\bf k}|^{\mu} + \cdots$
for small values of ${\bf k}$ and $\mu\le 2$, implying that the
random walks is non biased. In the long
time limit  
$P\left( {\bf k }, t_a , t \right) \sim  \exp( - m_{\mu} |k|^{\mu} t )$,
and either a L\'evy behavior ($\mu < 2)$ or a Gaussian 
behavior $(\mu=2$) is found, 
as expected from the Gauss--L\'evy central limit theorem \cite{Feller}.
In what follows, we investigate cases when this standard 
behavior does not hold.

\section{Asymptotic Behaviors} 
\label{SecAB}

 Let us now consider basic properties of ACTRWs. 
While Eq. (\ref{eq09}) is valid for a large class of
random walks, including
L\'evy flights $(\mu < 2)$, 
we will 
assume that variance
of $f({\bf r})$ is finite 
($\mu=2$). Special emphasis will be given to the
case when $\psi(t)$ is moment-less $\alpha<1$,
since this regime exhibits 
interesting aging behaviors. 

\subsection{Mean Square Displacement}
\label{SecMean}

By differentiating Eq. (\ref{eq09}) with respect to
${\bf k}$ and setting ${\bf k}=0$, we obtain the moments of the random 
walk in a standard way. 
Assuming a non biased symmetrical random walk, we obtain
\begin{equation}
\langle r^2 \left( s, u \right) \rangle = 
{ h_s(u) m_2 \over u \left[ 1 - \psi\left(u\right) \right] },
\label{eq11}
\end{equation}
where $m_2=\int r^2 f(|{\bf r}|) {\rm d} {\bf r}$ is assumed to be 
finite.
We consider power law waiting time PDFs
as in Eqs (\ref{eqpprr},\ref{eqpprr1}),
in the limit where both $u$ and $s$ are small, their ratio being
arbitrary, we find
\begin{equation}
\langle r^2 \left( s, u \right) \rangle \sim { \left( u^{\alpha} - s^{\alpha} \right) m_2 \over s^{\alpha} \left( u -s\right) A u^{1 + \alpha} } .
\label{eqx2ass}
\end{equation}
As shown below one can invert this equation exactly to the double time
domain. However it is instructive to consider
two limits first. 
If $u\ll s$, corresponding to $t \gg t_a$,
we have
\begin{equation}
\langle r^2 \left( s, u \right) \rangle \sim { u^{-1-\alpha}m_2\over A s}. 
\label{eqx2assa}
\end{equation}
While for $s \ll u$, corresponding to $t_a \gg t$, we have
\begin{equation}
\langle r^2 \left( s, u \right) \rangle \sim { m_2\over A u^2 s^{\alpha} } .
\label{eqx2ass1}
\end{equation}
Inverting Eq. (\ref{eqx2assa}) and Eq. (\ref{eqx2ass1}), we obtain
\begin{equation}
\langle r^2 \left( t_a , t \right) \rangle \sim \left\{
\begin{array}{l l}
{ m_2 t^{ \alpha} \over A \Gamma\left( 1+ \alpha \right) } \ & t \gg t_a \\
\ & \  \\
{ m_2 t t_a^{\alpha -1}  \over A \Gamma(\alpha) } \ & t \ll t_a. 
\end{array}
\right.
\label{eqxxxxe}
\end{equation}
This result is valid provided that both $t,t_a \gg A^{1/\alpha}$.
In the limit $t \gg t_a$ we recover standard
behavior found in non-equilibrium CTRW \cite{Weiss1}. In the aging regime,
$t \ll t_a$ we find an interesting behavior. Independent
of the exponent $\alpha$, the mean square displacement
increases linearly with respect to the forward
time $t$, as found in normal diffusion processes.
In addition, the
diffusion is slowed down as the age of the random walk $t_a$
is increased. This behavior is expected,
due to statistically longer release times,
from the initial position of the particle, as the age of the
random walk is increased.  

 We now consider a specific choice  of waiting time distribution:
\begin{equation}
\psi(u) = { 1\over 1 + A u^{\alpha} },
\label{eqppssii}
\end{equation}
corresponding to \cite{Hilfer}
\begin{equation}
\psi(t) = {t^{\alpha -1} \over A} E_{\alpha,\alpha}\left( - { t ^{\alpha}\over A} \right),
\label{eqHil}
\end{equation}
where $E_{\alpha,\alpha}(x)$ is the generalized Mittag--Leffler
function \cite{Erdelyi}. 
Inserting Eq. (\ref{eqppssii}) in Eq.
(\ref{eq11}) we have
\begin{equation}
\langle r^2 \left( s, u \right) \rangle = {m_2 \over A} { u^{\alpha} - s^{\alpha} \over s^{\alpha} \left( u - s \right) } { 1 \over u^{ 1 + \alpha} } .
\label{eqx2exac}
\end{equation}
Hence for this choice of waiting times,
 Eq. (\ref{eqx2ass}) is exact and not limited
to the asymptotic regime. Inverting to the
time domain using Eq. 
(\ref{eq05}) we find
\begin{equation}
\langle r^2 \left( t_a , t \right) \rangle = { m_2 \over A} { \sin \left(\pi \alpha \right) \over \pi } {1 \over t_a \left({ t \over t_a} \right)^{\alpha} \left( 1 + { t \over t_a} \right) } \otimes {t^{\alpha} \over \Gamma\left( 1 + \alpha \right) },
\label{eqxom}
\end{equation}
where $\otimes$ is the Laplace convolution operator
with respect to the forward time $t$.
We rewrite Eq. (\ref{eqxom}) as
\begin{equation}
\langle r^2 \left( t_a , t \right) \rangle = t_a^{\alpha} { m_2 \over A \Gamma(1 + \alpha)} { \sin \left(\pi \alpha \right) \over \pi } \int_0^{t/t_a} { \left( t/t_a - y \right)^{\alpha} \over y^{\alpha} \left( 1 + y \right) } {\rm d } y.
\label{eqxom1}
\end{equation}
The solution of the integral is readily obtained, we find
\begin{equation}
\langle r^2 \left( t_a , t \right) \rangle = { m_2 \over A} { 1 \over \Gamma\left( 1 + \alpha \right) } \left[ \left( t + t_a\right)^{\alpha} - t_a^{\alpha} \right] . 
\label{eqx2eexs}
\end{equation}
The right hand side  of Eq. (\ref{eqx2eexs}) describes the
long time $t$, long time $t_a$, behavior of a large class of
random walks with waiting time PDF satisfying
$\psi(t)\propto t^{- ( 1 + \alpha}) $ [i.e., since
the right hand side of Eq. (\ref{eqx2eexs}) is the double inverse Laplace
transform of Eq. (\ref{eqx2ass})].
Note that if $\alpha=1$ in Eq. (\ref{eqx2eexs}), the 
random walk does not exhibit aging.

\subsection{Green Function}

 In this subsection we
investigate asymptotic properties of the Green function $P({\bf r},t_a,t)$, 
by considering the continuum approximation
of Eq. ({\ref{eq09}). This approximation is expected
to work in the limit where both the forward time
$t$ and the aging time $t_a$ are long.  
A proof of the validity of this approach,
is given in Sec. \ref{Secproof}, for the one dimensional
ACTRW. 
We assume a symmetric random walk, hence for small $|{\bf k}|$, 
the following expansion is valid:
\begin{equation}
f({\bf k})\sim 1 - { 1 \over 2} |{\bf k}|^2 {m_2 \over d},  
\label{eqkexp}
\end{equation}
where $d$ is the dimensionality of the problem. 
We also use the small Laplace variable $u$ expansion
$\psi(u) \sim 1 - A u^{\alpha}$ in Eq. 
(\ref{eqpprr1}).
Inserting these expansions in Eq. (\ref{eq09}),
we obtain 
\begin{equation}
P\left( {\bf k}, u, s \right) \sim { s^{\alpha} u - s u^{\alpha} \over s^{\alpha + 1} u \left( u - s \right)} + {\left( u^{\alpha} - s^{\alpha} \right) \over s^{\alpha} \left(u - s\right) } { A u^{ \alpha - 1} \over A u^{\alpha} + |{\bf k}|^2 {m_2 \over  2 d}  }.
\label{eq13m1}
\end{equation}
For convenience, and without loss of generality, we choose
now to work in units where $A=1$ and $m_2/(2 d)=1$.

Inverting Eq. (\ref{eq13m1}) to the double time $(t_a,t)$
 -- real space ${\bf r}$  domain we find
$$ P({\bf r}, t_a , t) \sim 
p_{0} \left(t_a , t\right) \delta\left({\bf r} \right)  + $$
\begin{equation}
{ \sin\left( \pi \alpha \right) \over \pi } 
{ 1 \over  t_a \left( { t \over t_a} \right)^{\alpha} \left( 1 + { t \over t_a} \right) } 
\otimes P_{AMW}\left({\bf r}, t\right),
\label{eq13}
\end{equation}
where $\otimes$ in Eq. (\ref{eq13}) is the Laplace convolution operator
with respect to
the forward time $t$, and in this limit 
\begin{equation}
p_{0} (t_a , t) \sim { \sin\left( \pi \alpha\right) \over \pi } \int_{t/t_a}^{\infty} { {\rm d} x \over x^{\alpha} \left( 1 + x \right)}.
\label{eq14}
\end{equation}
In Eq. (\ref{eq13})
 $P_{AMW}({\bf r}, t)$ is the 
long time solution of the Montroll--Weiss equation,
i.e., the Green function of the fractional diffusion equation
\cite{review}, 
\begin{equation}
P_{AMW}({\bf r}, t) \equiv {\cal L}^{-1}  {\cal F}^{-1} \left\{  {  u^{  \alpha-1 } \over   u^{\alpha} + |{\bf k}|^2  } \right\}
\label{eq05a}
\end{equation}
where  ${\cal L}^{-1}  {\cal F}^{-1}$ is the inverse Laplace $u \to t$ 
inverse Fourier ${\bf k} \to {\bf r}$ operator.

The Green function Eq. (\ref{eq13}), is a sum of two terms.
The first term on the right hand side of Eq. (\ref{eq13}) is 
a singular term (i.e., the $\delta({\bf r})$ term). 
This term corresponds to random walks where number
of steps in time interval $(0,t)$ is zero. Unless
$t\gg t_a$, this term cannot be neglected, since without
it the Green function in Eq. (\ref{eq13}) is not normalized. 
Thus ACTRW exhibits a behavior different than ordinary
CTRWs, where realizations of random walks  where number of steps
is zero do not contribute to the asymptotic behavior.   

In one dimension, we have
\begin{equation}
P_{AMW}(r,t) = { t \over \alpha |x|^{1 + 2/\alpha} } l_{\alpha/2} \left( { t \over |x|^{2/\alpha} } \right),
\label{eq14p}
\end{equation}
where $l_{\alpha/2}(t)$ is the one sided L\'evy stable PDF, whose
Laplace pair is $\exp( - u^{\alpha/2} )$.
Hence the Green function solution of the ACTRW is
$$ P(x, t_a , t) \sim
p_{0} \left(t_a , t\right) \delta\left(x\right)  +  $$
\begin{equation}
{ \sin\left( \pi \alpha \right) \over \pi }
{ 1 \over  t_a \left( { t \over t_a} \right)^{\alpha} \left( 1 + { t \over t_a} \right) }
\otimes
{ t |x|^{-(1+2/\alpha)}  \over \alpha 2^{1/\alpha} } l_{\alpha/2} \left( { t |x|^{-(2/\alpha)}  \over  2^{1/\alpha}} \right).
\label{eq131d}
\end{equation}
Scaling Eq. (\ref{eq131d}) with $\tau \equiv t/t_a$ and
$q \equiv |x|/t^{\alpha/2}$ we obtain a scaled form of the Green
function:
$$ P(x, t_a , t) \sim
p_{0} \left(t_a , t\right) \delta\left(x\right)  +  $$
$$ {\sin\left( \pi \alpha \right) \over \pi \alpha} t^{ - \alpha/2} { 1 \over \tau q^{ 1 + 2 / \alpha } } \times  $$
\begin{equation}
\int_0^{\tau} {\rm d} \tau' { \tau' \over \left( \tau - \tau'\right)^{\alpha} \left( 1 + \tau - \tau' \right)} l_{\alpha/2} \left( {\tau'\over q^{2/\alpha} \tau } \right), 
\label{eq13aaaa}
\end{equation}
Below we analyze this expression in some detail.
%

In  $d$ space dimensions we have \cite{schneider}
$$ P_{AMW} \left( r , t \right) = \alpha^{-1} \pi^{-d/2} r^{ - d} \times $$
\begin{equation}
H_{12}^{20}
\left( 2^{- 2 /\alpha} r^{2 /\alpha} t^{-1} |
\begin{array}{l}
\left( 1 , 1 \right) \\
\left( d/2, 1/\alpha),(1,1/\alpha\right)
\end{array}
\right),
\label{eq23}
\end{equation}
Besides the $d=1$ case, the Fox function solution 
$H_{12}^{20}$
is not
generally tabulated, hence this solution is rather formal,
though asymptotic behaviors of Eq. (\ref{eq23}) are well
investigated \cite{schneider,Tran}. 
A practical method of obtaining the solution
of $P_{AMW}(r,t)$ Eq. (\ref{eq23}), 
using the inverse L\'evy transform,
is given in \cite{Tran}. 
Using this method, we find the integral representation
of the aging Green function
in $d$ dimension
$$ P(r, t_a , t) \sim
p_{0} \left(t_a , t\right) \delta\left({\bf r} \right)  +  $$
$$ { 1 \over \tau} { \sin \left( \pi \alpha \right) \over 2^d \pi^{d/2 + 1} \alpha } t^{ - \alpha d /2}  \times $$
\begin{equation}
\int_0^{\infty} {\rm d} s \int_0^{\tau} {\rm d } \tau' 
{ \left( \tau - \tau' \right) \over s^{1 + 1/\alpha + d/2} } 
l_{\alpha} \left[ { \left( \tau - \tau' \right) \over \tau s^{1/\alpha} } \right]
{ e^{- q^2 / (4 s )}  \over \tau'^{\alpha} \left( 1 + \tau' \right) },
\label{eqasss}
\end{equation}
where $q=r/t^{\alpha/2}$ and $\tau=t/t_a$.
Similar to the one dimensional case, this solution shows the precise relation
between ACTRWs and L\'evy stable laws.

To conclude, Eq. (\ref{eq13}) shows
that the asymptotic solution of ACTRW is a sum of
two terms. The first is  a singular term, and
the second is a convolution of the distribution of
the first waiting time
and the asymptotic Green function of the non-equilibrium
CTRW. 

\subsection{Graphic Examples}

In order to better understand the asymptotic behavior of the ACTRW,
we perform numerical calculations to obtain the non-singular
part of the Green function for different values of \( \alpha  \)
in one dimension. The one sided L\'evy stable probability density in
Eq.(\ref{eq14p}) was obtained using a numerical inverse Laplace transformation
method
\cite{DLM99,DLM99b}. The calculated PDF was then used to evaluate
the convolution integral numerically to obtain the non-singular part
of the Green function according to Eq.(\ref{eq13aaaa}).
For mathematical details
on one sided L\'evy stable laws,
see Appendix in \cite{Tran}, and references
therein.

In Figs. \ref{fig2}-\ref{fig5}, we present the calculated non-singular part of the Green
function for different \( \alpha  \). Fig. \ref{fig2} shows
a three dimensional plot of the scaled non-singular Green function,
\( P(q,\tau )\cdot t^{\alpha /2} \), versus scaling variables, \( \tau =t/t_{a}
\)
and \( q=|x|/t^{\alpha /2} \), for \( \alpha =1/2 \). A smooth transition
from the aging behavior when \( \tau \ll 1 \) to the non-equilibrium
CTRW behavior when \( \tau \gg 1 \) can be clearly seen in this figure,
where the time ratio \( \tau =t/t_{a} \) is changed continuously.
In addition, we observe a monotonic increase of the non-singular
Green function as the scaled time \( \tau  \) increases.
Note that as the scaled $\tau$ is increased the singular term
is decreasing, hence we may think of this aging process, as
if the singular part of the Green function, is feeding
the non-singular part. 

 Figs. 
\ref{fig3}-\ref{fig5} 
show the scaled non-singular Green function versus 
\( q=|x|/t^{\alpha
/2} \)
for 
\( \alpha =1/6 \), \( \alpha =1/2 \), and \( \alpha =5/6 \),
respectively. In each figure, the scaled Green functions
at several different \( \tau  \)s are shown.
A few general features can be seen in these figures. First,
the Green function is clearly non-Gaussian for all cases, as we expected.
A comparison between the shape of the Green function for \( \alpha =1/6 \)
in Fig. \ref{fig3} and for \( \alpha =5/6 \) in Fig. \ref{fig5} clearly demonstrates
that the deviations from Gaussian behavior are stronger for smaller
\( \alpha  \). In the limit of \( \alpha \rightarrow 1 \) (not shown),
we obtain a Gaussian Green function. Second, for \( t\gg t_{a} \)
(i.e. \( \tau \gg 1 \)), we recover the usual non-equilibrium CTRW
behavior shown as the solid curves in Figs. \ref{fig2}-\ref{fig5}. Finally, as \( \alpha  \)
becomes small, the convergence towards the non-aging behavior when
\( t/t_{a}=\infty  \) becomes extremely slow. For example, the case
for \( \alpha =1/6 \) in Fig. \ref{fig3} shows a significant deviation from
the non-equilibrium CTRW behavior (solid line) when \( t/t_{a}=10000 \).
As a result, the Green function exhibits aging behavior even when
\( t/t_{a} \) is large, and an ACTRW treatment for dynamics in this
regime will be essential.

%
%
%

\begin{figure}
\begin{center}
\epsfxsize=70mm
\epsfbox{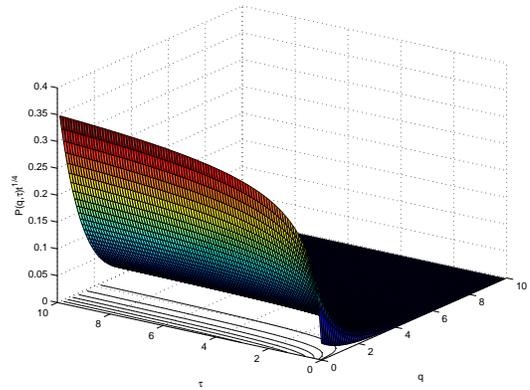}
\end{center}
\caption
{
Three dimensional plot of the scaled non-singular Green function
versus $\tau=t/t_a$ and $q=|x|/t^{\alpha/2}$ for $\alpha=1/2$ and one
dimension. 
 Notice a smooth transition from aging behavior $\tau<<1$ to 
Montroll-Weiss CTRW behavior found when $\tau>>1$.
}
\label{fig2}
\end{figure}

%

%
\begin{figure}[htb]
\epsfxsize=20\baselineskip
\centerline{\vbox{
	\epsffile{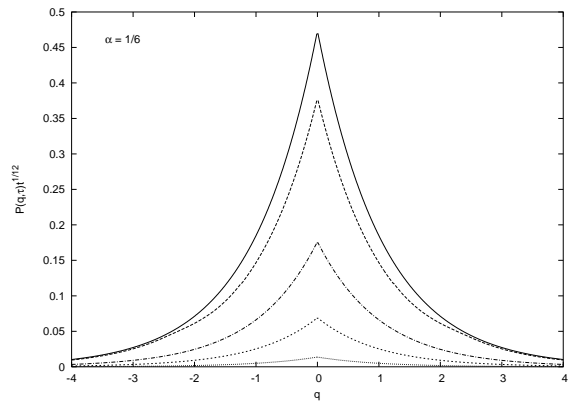}  }}
\caption {
The scaled non-singular Green function 
versus  $q= |x|/t^{\alpha/2}$
for $\alpha=1/6$ and for 
five different
$\tau=t/t_a$:
$\tau=0.1$ (dot-dash), $\tau=1$ (dash),
$\tau=10$ (dot dash), $\tau=10000$ (dashed), and $\tau=\infty$ (solid).
The solid curve is the asymptotic behavior of the
non-equilibrium CTRW.
Notice the non-Gaussian shape of the Green function and the 
slow convergence towards the non-aging CTRW behavior (the solid line),
compared with the cases $\alpha=1/2$ and $\alpha=5/6$ shown below.
}
\label{fig3}
\end{figure}

\begin{figure}[htb]
\epsfxsize=20\baselineskip
\centerline{\vbox{
	\epsffile{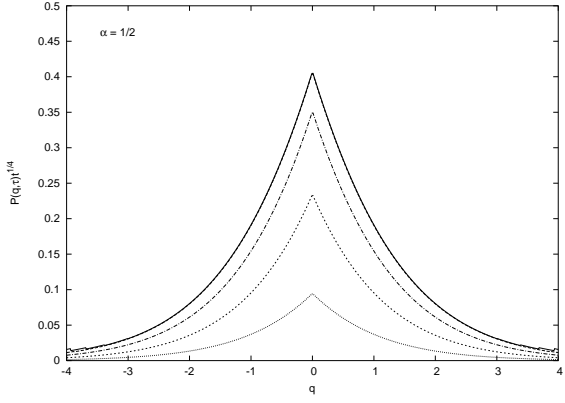}  }}
\caption {
Same as Fig. \protect{\ref{fig3}} for  $\alpha=1/2$. 
}
\label{fig4}
\end{figure}

\begin{figure}[htb]
\epsfxsize=20\baselineskip
\centerline{\vbox{
	\epsffile{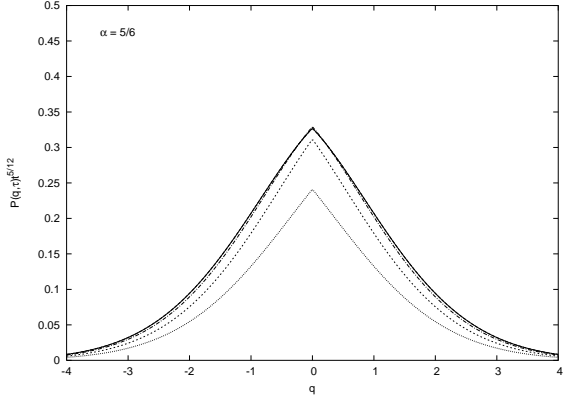}  }}
\caption {
Same as Fig. \protect{\ref{fig3}} for  $\alpha=5/6$.
}
\label{fig5}
\end{figure}

\subsection{Proof of Asymptotic Behavior for One Dimension}
\label{Secproof}

 We now prove the validity of Eq.  (\ref{eq13}) using a method
developed in \cite{Barkai}. The main idea is to show
that moments of the ACTRW, are in the asymptotic limit described
well by Eq. (\ref{eq13}).  For simplicity we assume a one dimensional
symmetric random walk. 

The moment generating function
$f({\bf k})$ is expanded 
\begin{equation}
f({\bf k} ) = 1 - m_2 {k^2 \over 2} + m_4 { k^4 \over 24} - m_6 {k^6 \over 720} + \cdots.
\label{eqeexx}
\end{equation}
Where $m_2$, $m_4$ etc are the moments of the  
jumps. Inserting this expansion in Eq. 
(\ref{eq09}) we obtain the small $k$  expansion
of the ACTRW moment generating function:
$$ P(k,s,u) = {1 \over su} - {h_s(u) \over u} \left\{ \left[ 1 + \Omega(u) \right] m_2 { k^2 \over 2}  \right. $$
$$ -\left[ 1 + \Omega(u) \right] \left( m_4 + 6 \Omega(u) m_2^2 \right) { k^4 \over 24} $$
\begin{equation}
\left[ 1 + \Omega(u) \right] \left[ m_6 - 30 \Omega(u) m_2 m_4 + 90 \Omega^2(u) m_2^3 \right] { k^6 \over 720} + \cdots   
\label{eqeexxa}
\end{equation}
where
\begin{equation}
 \Omega(u) = \psi(u)  / [1 - \psi(u)].
\label{eqOM}
\end{equation}
The moments $\langle x^n (s,u) \rangle$
of the ACTRW are defined in the usual way
\begin{equation}
P(k,s,u) = \sum_{n=0}^{\infty} \langle x^n (s,u) \rangle { ( i k)^n \over n!}.
\label{eqMGF}
\end{equation}
Comparing Eq. (\ref{eqeexxa})
with Eq. (\ref{eqMGF}) we have
\begin{equation}
\langle x^0 \rangle = { 1 \over s u},
\label{eqNor}
\end{equation}
which means that normalization is conserved.
For the second moment we obtain
\begin{equation}
\langle x^2 ( s , u ) \rangle = { h_s (u ) \over u} \left[ 1 + \Omega(u) \right]
\label{eqx2}
\end{equation}
which is the same as Eq. 
(\ref{eq11}). The fourth moment is more interesting:
\begin{equation}
\langle x^4 ( s, u ) \rangle = { h_s(u) \over u} \left[ 1 + \Omega(u) \right]
\left[ m_4 + 6 \Omega(u) m_2^2 \right].
\label{eqx4}
\end{equation}
Higher order moments are obtained in a similar way, for the sake of space they
are not included here. Odd moments vanish due to the symmetry
of the random walk. 

 One can easily see that the ACTRW $n$ th  moment $\langle x^n ( s, u) \rangle$
depends on the microscopic jump moments $\{m_2, \cdots, m_n\}$. However
in the limit $u \to 0$
\begin{equation}
\langle x^n ( s, u) \rangle \sim {h_s(u) \over u} {1 \over ( A u^{\alpha})^{n/2}} m_2^{n/2} {n! \over 2^{n/2}},
\label{eqASmom}
\end{equation}
which depends on $m_2$ but not on the higher order jump moments $m_4,m_6$ etc.
Thus the moments $m_n$ with $n> 2$ are the irrelevant parameters
 in this problem.
Inserting Eq. (\ref{eqASmom}) in Eq. 
(\ref{eqMGF}) we have
\begin{equation}
P(k,u,s) \sim \sum_{n=0}^{\infty} {h_s(u) \over u} \left( i k \right)^n \left( {m_2 \over 2 A u^{\alpha} }\right)^{n/2}.
\label{eqddd}
\end{equation}
Since we are interested in the limit where $t$ and $t_a$ are large, the 
ratio $t/t_a$ being arbitrary, the corresponding Laplace variables $u$ and $s$
must approach zero their ratio being arbitrary. Therefore $h_s(u)$ in Eq.
(\ref{eqddd}) is given by its asymptotic form in Eq. (\ref{eq06}).
Inserting this expression for $h_s(u)$ into Eq. (\ref{eqddd}), setting $m_2=1$,
and then summing over $n$, we find an expression
for $P(k,u,s)$ that is the same as Eq. (\ref{eq13m1}).
Eq. (\ref{eq13m1}) when transformed
yields Eq. (\ref{eq13}).
To conclude we showed that Eq. (\ref{eq13m1})
describes the small $s,u$ behavior of the ACTRW moments, hence it 
follows that Eq. (\ref{eq13}) describes the long time $t$ and $t_a$
behavior of the ACTRW Green function, the set of moments $m_4,m_6$ etc
are unimportant in this limit.

\subsection{Behavior on the Origin}

Using Eq. (\ref{eq14p}) we investigate the nonsingular part of
the ACTRW on the origin. For $d=1$,
we find
\begin{equation}
P(x,t_a,t)|_{x=0} = t^{ - \alpha/2} g\left( { t \over t_a} \right),
\label{eq15}
\end{equation}
where 
\begin{equation}
g(z) = z^{ \alpha /2} { \sin\left( \pi \alpha \right) \over 2 \pi \Gamma\left( 1 - \alpha/2\right)} \int_0^z {\rm  d} y { \left( z - y \right)^{ - \alpha/ 2} \over \left( 1 + y \right) y^{ \alpha} } .
\label{eq16}
\end{equation}
Hence 
\begin{equation}
P(x,t_a,t)|_{x=0} \sim  \left\{
\begin{array}{l l}
{ t^{ - \alpha/2} \over 2 \Gamma\left( \alpha \right) \Gamma \left( 2 - 3 \alpha /2 \right) } \left( { t \over t_a } \right)^{ 1 - \alpha} \ & t \ll t_a \\
\ & \  \\
{ t^{ - \alpha/2} \over 2 \Gamma\left( 1- \alpha/2 \right) } \ & t \gg t_a .
\end{array}
\right.
\label{eq17}
\end{equation}
In the limit $t>>t_a$ we recover standard CTRW behavior \cite{Weiss1}.

In Fig. \ref{fig6}, we present the behavior of the ACTRW on the origin.
The ratio of the scaled non-singular ACTRW Green function to the
Montroll-Weiss non-equilibrium CTRW Green function on the origin,
$P(x,t,t_a)t^{\alpha/2}|_{x=0} / 2 \Gamma( 1- \alpha/2)$,
is plotted versus the scaled dimensionless time $t/t_a$. This ratio
approaches one in the limit $t>>t_a$, showing that the ACTRW process
will converge to the standard non-equilibrium CTRW behavior when $t>>t_a$.
It can be clearly seen in the Fig. \ref{fig6} that for $\alpha > 1/2$,
the ACTRW process has roughly converged
to the non-equilibrium CTRW limit when
$t/t_a \simeq 1$,
while for $\alpha \to 0$, the crossover to CTRW limit becomes extremely slow.
For example, when $\alpha=1/12$,
large deviations from the CTRW limit
are clearly observed even when $t/t_a=10^{8}$.
Since the limit $\alpha \to 0$ is important for several
systems \cite{Parisi,Fisher,KlafterD},
it becomes clear that when $\alpha$ is small, the convergence towards
the standard CTRW results becomes extremely slow, and
the aging effect is of importance even when $t>t_a$.

\begin{figure}[htb]
\epsfxsize=20\baselineskip
\centerline{\vbox{
	\epsffile{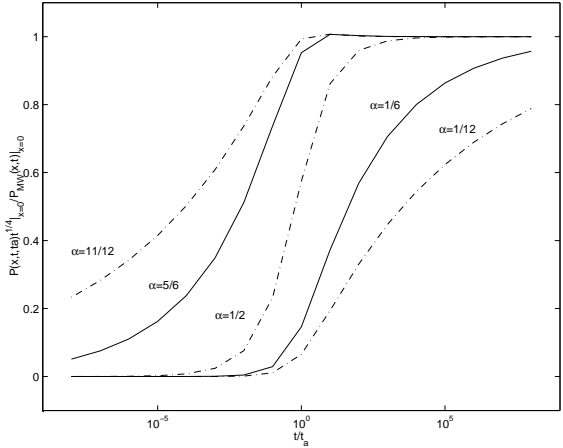}  }}
\caption {
The behavior of the non-singular
 Green function on the origin normalized
by the long time solution of the Montroll--Weiss non-equilibrium CTRW. 
The convergence of ACTRW towards the CTRW result
is extremely slow when $\alpha \to 0$.
}
\label{fig6}
\end{figure}

From Eq. (\ref{eq17}),
we see that for $t \ll t_a$ the non-singular part of $P(x,t_a,t)|_{x=0}$
increases with time $t$ when $\alpha<2/3$. This unusual behavior
is not unphysical, because the singular delta function term
is a decreasing function of
time, and the total probability of finding the random walker in a
small vicinity of the origin is decreasing monotonically with
time as expected.

\section{Biased ACTRW}
\label{SecBi}

 We now consider one dimensional biased ACTRW. We therefore
use the small $k$ expansion of $f(k)$:
\begin{equation}
f(k) = 1 + i k m_1 - { k^2 \over 2} m_2 \cdots,
\label{eqBi01}
\end{equation}
where $m_1>0$ is the averaged jump length. 
Differentiating Eq. (\ref{eq09}) once with respect to $k$
and taking $k=0$ we find the mean displacement of the
random walker in $s,u$ space:
\begin{equation}
\langle x\left( s, u \right) \rangle = { m_1 h_s \left( u \right) \over u \left[ 1 - \psi\left( u \right) \right]  },
\label{eqBi02}
\end{equation}
where $h_s(u)$ is defined in Eq. (\ref{eq04}).
Differentiating Eq. (\ref{eq09}) twice with respect to $k$,
we find the second moment of the biased random walk
\begin{equation}
\langle x^2 \left( s , u \right) \rangle = { h_s \left( u \right) \over u \left[ 1 - \psi\left( u \right) \right]} \left[ 2 m_1^2 { \psi\left( u \right) \over 1 - \psi \left( u \right) } + m_2\right].
\label{eqBi03}
\end{equation}

\subsubsection{Aging Einstein Relation}

 We now derive a relation between the mean square displacement
in the absence of bias, and the mean displacement of the
particle in the presence of bias, reflecting the fluctuation--dissipation
relation valid within linear response theory (see \cite{FDT1,FDT2,FDT3,FDT4}
for related work). The case of zero
aging $t_a=0$ were discussed in \cite{BarkaiEin}, where some conceptual
problems of linear response theory for systems exhibiting anomalous
type of diffusion was discussed (e.g., the non-stationarity of the process,
the dependence of $\alpha$ on external field). 

 We assume that the random walk is on a one dimensional
lattice with lattice spacing $a$, therefore
\begin{equation}
f(x) = P_L \delta ( x - a) + P_R \delta(x+a). 
\label{eqBi04} 
\end{equation}
Here $P_L + P_R = 1$, hence the jump moments
in Eq.
(\ref{eqBi01}) are
 $m_1= ( P_R -  P_L ) a$ and 
$m_2 = a^2$. We assume that the process obeys local detailed balance,
namely $P_L / P_R = \exp \left( - a F / k_b T \right)$ where $T$ is the 
temperature. Using these conditions, and the assumption of
weak field  $a F / k_b T \ll 1$, we
have $m_1\simeq a^2 F / 2 k_b T $. Using Eqs.  
(\ref{eq11}) and
(\ref{eqBi02}), we find
\begin{equation}
\langle x \left(s,u \right) \rangle_F  = { F \over 2 k_b T } \langle x^2 \left( s , u \right) \rangle_0.
\label{eqBi04a}
\end{equation}
The subscript $F$ in Eq. (\ref{eqBi04a}) indicates the presence of
external field $F$. $\langle x^2 \left(s , u \right) \rangle_0$ is the
mean square displacement in the absence of a field, i.e. Eq. (\ref{eqBi03})
with $m_1=0$ and $m_2=a^2$. 
Since the equation holds for the $s,u$ domain, it holds
also for the $t_a,t$ domain
\begin{equation}
\langle x\left( t_a , t \right) \rangle_F = { F \over 2 k_b T } 
\langle x^2\left( t_a , t \right) \rangle_0  .
\end{equation}
 Thus the mean square displacement of the
particle in the absence of the field (the fluctuation) yields the
mean displacement in the presence of a weak field. When 
the waiting times are exponentially distributed,
we obtain the usual Einstein relation between mobility and diffusion constant,
which is independent of the age of the process $t_a$.
For experimental verification of Eq. (\ref{eqBi04a}) in the non-aging
regime $t_a=0$ and with $\alpha<1$, see \cite{Qu,Amb,Com}.

\subsubsection{Asymptotic Behavior of Biased ACTRW}

From Eq. (\ref{eqBi02}) we can derive the behavior of the mean
displacement in exactly the same way as done in Sec.
\ref{SecMean}, and find
\begin{equation}
\langle x \left( t_a , t \right) \rangle \sim \left\{
\begin{array}{l l}
{ m_1 t^{ \alpha} \over A \Gamma\left( 1+ \alpha \right) } \ & t \gg t_a \\
\ & \  \\
{ m_1 t t_a^{\alpha -1}  \over A \Gamma(\alpha) } \ & t \ll t_a. 
\end{array}
\right.
\label{eqBi05}
\end{equation}
For the second moment we use the small $s,u$ behavior of
Eq. \ref{eqBi03} and find
\begin{equation}
\langle x^2 \left( s , u \right) \rangle \sim { 1 \over A u^{ 1 + \alpha} }
{ u^\alpha - s^{ \alpha} \over s^{\alpha} \left( u - s \right) } \left[
{2 m_1^2  \over u^\alpha A} + m_2 \right].
\label{eqBi06}
\end{equation}
Inverting this equation using Eq.  
(\ref{eqBi05}), we investigate now the dispersion:
\begin{equation}
\sigma\left(t_a, t \right) \sim \sqrt{ \langle x^2 \left( t_a, t \right) \rangle - \langle x\left( t_a , t \right) \rangle^2 }. 
\label{eqBi07}
\end{equation}

Considering first the $u\ll s$ limit corresponding to
$t \gg t_a$ we recover Shlesinger's result \cite{Shlesinger}
$$ \sigma\left(t_a, t \right) \sim  $$
\begin{equation}
\sqrt{{ m_1^2 t^{ 2 \alpha} \over A^2}
\left[ { 2 \over \Gamma\left( 1 + 2 \alpha\right) } - {1 \over \Gamma^2\left( 1 + \alpha\right)} \right] + { m_2 t^{\alpha} \over A \Gamma \left( 1 + \alpha \right) } },
\label{eqBi07a}
\end{equation}
hence if $m_1 \ne 0$ and $\alpha < 1$, one finds
\begin{equation}
\sigma\left(t_a, t \right) \sim { m_1 t^\alpha \over A} \sqrt{
 { 2 \over \Gamma\left( 1 + 2 \alpha\right) } - {1 \over \Gamma^2\left( 1 + \alpha\right)}}. 
\label{eqBi08}
\end{equation}
For $t\gg t_a$ the dispersion of the biased CTRW grows like the mean
Eq. (\ref{eqBi05}),
a behavior very different than normal Gaussian diffusion.

Considering the $s\ll u$ limit of Eq. (\ref{eqBi06}) and using Eq.
(\ref{eqBi05}), 
we find for aging limit $t \ll t_a$
\begin{equation}
\sigma\left(t_a, t \right) \sim 
\sqrt{
{ m_2 t t_a^{\alpha - 1} \over A \Gamma\left( \alpha \right) } + 
{ 2 m_1^2 \over A^2} { t_a^{\alpha - 1} t^{ 1 + \alpha} \over \Gamma\left( \alpha \right) \Gamma\left( 2 + \alpha \right) } - { m_1^2 t^2 t_a^{2 \alpha - 2} \over A^2 \Gamma^2\left( \alpha \right)}.
}
\label{eqBi09}
\end{equation}
Hence if $m_1 \ne 0$ and $\alpha < 1$, one finds
$$ \sigma\left(t_a, t \right) 
\sim  $$
\begin{equation}
{ m_1 \over A} \sqrt{ { 2 \over \Gamma\left( \alpha \right) \Gamma\left( 2 + \alpha \right) } } { t^{{1 + \alpha \over 2}} \over t_a^{ {1 - \alpha \over 2} } },
\label{eqBi10}
\end{equation}
which is valid for $t,t_a \gg A^{1/\alpha}$ and $t \ll t_a$.
As expected the dispersion decreases as
age of the processes becomes older. Note that as $\alpha \to
0$, the first term in eq. (\ref{eqBi09}) becomes important.

 For $\alpha=1$ one finds
\begin{equation}
\sigma \left( t \right) \sim \sqrt{ m_2 t \over A}, 
\label{eqBi11}
\end{equation}
where in this case $A$ has the meaning of the mean time between jumps.
The dispersion in this case is independent of the age of the
system $t_a$, as expected from normal diffusion.
We see that the dispersion in normal diffusion processes
is controlled by the second moment of jump lengths $m_2$ (even
when $m_1 \ne 0$), while for CTRW and ACTRW 
$m_1 \ne 0$  is the relevant parameter.

\section{Possible Application: Scher-Montroll Transport}

In this section we briefly point out to one possible
application of ACTRW.
Scher and Montroll \cite{SM} modeled transport in disordered
medium based on CTRW theory.
The fundamental
reasons of why and when their approach is valid,
while being the subject of much theoretical research \cite{SP,LV},
are not totally solved. 
 What is clear is that on a phenomenological level,
one can use the Scher--Montroll approach to
fit behaviors of charge currents
in a large  number of experiments in very different systems. 
For example, transport of charge carriers
in:
organic photo-refractive glasses \cite{h22}, nano-crystalline $T_i O_2$
electrodes \cite{Nelson},  conjugated polymer system poly p-phenylene 
\cite{poly,Pandey}, and liquid crystalline xinc octakis \cite{liqCry}.

 Scher and Montroll model such transport processes using 
{\em non-equilibrium} biased CTRW theory with
an effective waiting time distribution. 
In experiments, this corresponds to charge transport which is
started at time $t=0$, for example by a short photo flash 
applied on the system.
After the initial triggering of the process,
the charge carriers are transported
using an external bias. 
For such initial conditions,
we know that it is useful to assume that  the physical
transport process is described by the non-equilibrium
biased CTRW. 

 In an aging experiment one would start the process, by an external
impulse (e.g., a photo flash), then wait for an aging period $t_a$, and
only after that period add the external bias. 
In this case, biased aging CTRW might become a useful
tool describing the aging transport. 
At this time it is still an
open question if ACTRWs can be used to describe aging in real
systems. 
 Further it is not clear if aging in the above
mentioned systems \cite{h22,Nelson,poly,Pandey,liqCry} 
is measurable, and if so do these 
very different systems exhibit any common aging effects in their
transport?

In the non-aging experiments
\cite{h22,Nelson,poly,Pandey,liqCry} 
 one measures the current of charge carriers,  
which according to the predictions of Scher and Montroll 
exhibit a universal behavior: $I(t) \propto t^{\alpha- 1}$ for short times and
 $I(t) \propto t^{-\alpha- 1}$ for long times.
The transition time $t_L$
between these two behaviors depends among other things
on the length of the system.  The short time
behavior corresponds to 
$I(t) \propto {{\rm d} \over {\rm d} t} \langle x \rangle$,
with the non-equilibrium CTRW behavior $\langle  x \rangle \propto t^{\alpha}$,
which yields immediately $I(t) \propto t^{\alpha- 1}$ .
The long time behavior is more complicated,
and is due to absorbing boundary condition.

In this paper 
we have calculated the mean displacement of the biased ACTRW, without
including the influence of the boundary. Thus we provide 
the aging corrections to the short time behavior of  
Scher--Montroll transport. 
According to Eq. (\ref{eqBi05}), the Scher--Montroll
behavior $I(t) \propto t^{\alpha- 1}$
is replaced with $I(t) \propto t_a^{\alpha - 1}$ when $t \ll t_a \ll t_L$.
This behavior is independent of $t$, similar to behavior of
normal currents. In this aging regime,
the current decreases as $t_a$ is increased,
while in the non-aging case the current is decreasing when the
forward time $t$
is increased.
The assumption made
is that the dispersion of the probability packet
during the aging period, is small compared with
the length of the system. A detailed investigation of aging in Scher-Montroll
transport systems will be given elsewhere.  

\section{Summary}

 We have derived an exact expression for the Green function
of ACTRW in Fourier--double Laplace space. This generalized Montroll
-- Weiss equation describes dynamics of a large class of random walks.
Since the CTRW describes a large class of physical and chemical systems,
mainly disordered systems, we expect that ACTRW will be a valuable tool
when aging effects are investigated in these systems.
Interesting aging behaviors are found when the system turns non-ergodic,
namely when the mean waiting time diverges, $\alpha<1$.
We note that also
when the mean waiting time is finite, aging behaviors may be observed,
however only
within a certain time window. 

 We showed that asymptotic behavior of the Green function is related to
a few parameters of the underlying walk $\alpha,A$, and $m_2$, while other
informations contained in $\psi(t)$ and $f({\bf r})$ are irrelevant. 
The Green function behavior is non Gaussian when $\alpha<1$, it is 
related to L\'evy's generalized central limit theorem and
to Gordeche--Luck's \cite{GL} fractal renewal theory. Unlike standard
random walks or non-equilibrium  CTRWs, the asymptotic Green function
is a sum of two terms: a singular term corresponding to
random walks where number of jumps is zero and a non-singular term 
corresponding to random walks where number of jumps is one or more.
Finally, we note that the fractional Fokker-Planck equation framework
\cite{review,MBK},
developed based on CTRW concepts \cite{barkai9}, 
can be modified based on the
results obtained in this manuscript to include aging effects.
This topic is left for a future publication.

$$ $$
{\bf Acknowledgments} EB thank J. P. Bouchaud for pointing out Ref.
\cite{GL}. Discussions with L. Levitov and A. Heuer motivated this work.
 
\section{Appendix A}

 In this Appendix, we derive  Eq.
(\ref{eq04}) using a method which is slightly different then the
one used in \cite{GL}. Consider a non-equilibrium renewal
process which starts at time $t=0$ (in the ACTRW this initial
time is $-t_a$). Let
${\bar t}_i$ $i=1,...N$ be dots on the time axis
on which jumping events occur. Let ${\bar t}_{N+1}$ denote the
time on which the first jump event occurred which is larger than
$t_a$, namely $\bar{t}_{N} < t_a < \bar{t}_{N+1}$. Note that
$N$ itself is not fixed, it is a random variable.
The time intervals between jumps events are denoted by 
$\tau_i\equiv {\bar t}_{i+1} - {\bar t}_i$.

 The random variable we are interested in is $t_1$, where
$t_1\equiv \bar{t}_{N+1} - t_a$. Since $t_a$ is a parameter 
in the problem, knowledge of statistical properties of
$\bar{t}_{N+1}$ yields the distribution of $t_1$.
Hence let $P_{t_a} ( \bar{t}_{N+1})$ denote the PDF of the random
variable  $\bar{t}_{N+1}$.
It is given by:
$$ P_{t_a} \left(\bar{t}_{N+1} \right)  $$
\begin{equation}
= \sum_{N=0}^{\infty} \langle \delta\left( \bar{t}_{N+1} - \sum_{i=1}^{N+1} \tau_i\right) I \left( \bar{t}_N< t_a < \bar{t}_{N+1} \right) \rangle_N.
\label{eqA01}
\end{equation}
Here $I(\bar{t}_N < t_a < \bar{t}_{N+1})=1$ if $\bar{t}_N < t_a < \bar{t}_{N+1}$, otherwise it is
zero. The average in Eq. (\ref{eqA01}) is
\begin{equation}
\langle \cdots \rangle_N = \langle \Pi_{i=1}^{N+1} \int_0^{\infty} \psi(\tau_i) {\rm d} \tau_i \cdots \rangle.   
\label{eqA02}
\end{equation}

 We consider the double Laplace transform $t_a \to s$ and ${\bar t}_{N+1} \to u$
of $P_{t_a} \left( {\bar t}_{N+1} \right)$ Eq. (\ref{eqA01}),
$P_s(u)$. Using  ${\bar t_N} = \sum_{i=1}^N {\tau_i}$, we
have
$$\int_0^\infty e^{ - t_a s} I \left( {\bar t}_N < t_a < {\bar t}_{N+1} \right){\rm d} t_a  
=$$
\begin{equation}
{ e^{- {\bar t}_N s} - e^{ - {\bar t}_{N+1} s } \over s}=
{ e^{- s \sum^N_{i=1}  \tau_i } - e^{ -s \sum_{i=1}^{N+1} \tau_i } \over s},
\label{eqA03}
\end{equation}
and
\begin{equation}
\int_0^{\infty} e^{ - u \bar{t}_{N+1} } \delta\left( {\bar t}_{N+1} - \sum_{i=1}^{N+1} \tau_i \right) {\rm d} {\bar t}_{N+1} = e^{ - u \sum_{i=1}^{N+1} \tau_i}.
\label{eqA04}
\end{equation}
Using Eqs. (\ref{eqA03},\ref{eqA04}), 
we have from Eq. (\ref{eqA01})
$$ P_s(u) = \sum_{N=0}^{\infty} $$
\begin{equation}
\langle \exp\left( - u \sum_{i=1}^{N+1} \tau_i  \right)  { \exp\left(- s\sum_{i=1}^N \tau_i   \right) - \exp \left( - s \sum_{i=1}^{N+1} \tau_i \right) \over s} \rangle. 
\label{eqA05}
\end{equation}
Using the fact that the random variables
 $\tau_i$ are independent and identically
distributed, we have
\begin{equation}
P_s(u) = {1 \over s} \sum_{N=0}^{\infty} \left[ \psi^N(u+s)\psi(u) - \psi^{N+1}(u+s) \right],
\label{eqA06}
\end{equation}
where
$\psi(u+s)= \int_0^{\infty} \exp[ - (u + s) \tau] \psi( \tau) {\rm d} \tau$.  
Summing Eq. (\ref{eqA06}) we find
\begin{equation}
P_s(u) = { \psi(u) - \psi(u+s) \over s} { 1 \over 1 - \psi(u+s) }.
\label{eqA07}
\end{equation}

 Now the PDF $h_{t_a}(t_1)$ is obtained from $P_{t_a}(\bar{t}_{N+1})$
using $t_1={\bar t}_{N+1} - t_a$. According to definition of Laplace
transform, we can write
 $$ h_{t_a}(t_1)= $$
\begin{equation}
 {\cal L}^{-1} _{u \to t_1, s \to t_a} \left\{
\int_0^{\infty}  {\rm d} t_a  e^{- s t_a} \int_0^{\infty} d t_1 e^{- u t_1} h_{t_a} (t_1) \right\}, 
\label{eqA08}
\end{equation}
where ${\cal L}^{-1}$ is the double inverse Laplace transform.
We use $t_1=\bar{t}_{N+1} - t_a$ and find:
 $$ h_{t_a}(t_1) = $$ 
\begin{equation}
{\cal L}^{-1} _{u \to t_1, s \to t_a} \left\{
\int_0^{\infty}  {\rm d} t_a  e^{- s t_a} \int_{t_a}^{\infty} d t_1 e^{ -u ({\bar t}_{N+1} - t_a)} P_{t_a} ({\bar t}_{N+1}) \right\}. 
\label{eqA08a}
\end{equation}
Hence it is easy to see that
\begin{equation}
h_s(u) = P_{s-u} (u).
\label{eqA08b}
\end{equation}
Inserting Eq. (\ref{eqA07}) in Eq. (\ref{eqA08b}), we find
Eq. (\ref{eq04}).


\begin{thebibliography}{99}
\bibitem{remark} The main mathematical \cite{Feller} tool in investigating
asymptotic behavior of ACTRW will be the Tauberian theorem:
if $g(u) \sim u^{- \theta}$ as $u \to 0$, with $\theta > 0$ then
$g(t)\sim t^{\theta - 1}/\Gamma\left( \theta \right)$ as $t \to \infty$
where
$g(u)$ and $g(t)$ are Laplace pairs.
\end{thebibliography}
\end{document}